\documentstyle[preprint,aps]{revtex}
\tighten
\begin{document}

\maketitle
\noindent 
{\bf COMMENT ON "QUANTUM PHASE SHIFT CAUSED BY SPATIAL\\
CONFINEMENT"}
 \\
\\
\\
\indent\indent\indent\indent\indent{\bf Murray Peshkin} \\
\\
\indent\indent\indent\indent\indent {\it Argonne National Laboratory} \\ 
\indent\indent\indent\indent\indent {\it Physics Division-203} \\
\indent\indent\indent\indent\indent {\it Argonne, IL 60439-4843} \\
\indent\indent\indent\indent\indent {\it peshkin@anl.gov} \\
\\
\indent\indent\indent\indent\indent Received  3 September 1999; revised 22 November 1999 \\
\\
\\
\noindent
The analysis of phase shifts in executed and proposed interferometry
experiments on photons and neutrons neglected forces exerted at the
boundaries of spatial constrictions.  When those forces are included it
is seen that the observed phenomena are not in fact geometric in nature.
A new proposal in the reply to this comment avoids that pitfall.
\\
\\
\noindent
{\bf 1. INTRODUCTION} \\
\\
\noindent
Allman {\it et al.} [1] have proposed a neutron interference experiment 
wherein the neutrons in one arm of an interferometer pass through a 
channel in an otherwise reflecting barrier and the resulting phase 
shift is to be measured.  Following an earlier discussion [2], they 
calculate that the phase shift expected to be induced by the neutrons' 
passage through the channel in the barrier will be given for an 
appropriate range of the parameters by

\begin{equation}
\Delta \Phi \approx {\pi \over 4}\,\, {\lambda \ell \over a^2} =
{\pi ^2 \hbar \ell \over 2a ^2 \sqrt{2mE}} \,\,, 
\end{equation}
where $\ell$ and $a$ are the length and width of the channel 
and $\lambda$, $m$, and $E$ are the wavelength, mass, and energy of 
the neutrons.  They assert that no force is exerted on the neutrons 
and from that they conclude that the proposed experiment will 
demonstrate a new, purely geometrical, force-free effect of the 
Aharonov-Bohm type.

That conclusion is not correct.  The phase shifts induced by 
force-free interactions are necessarily independent of the energy 
of the neutrons [3,4,5], contrary to Eq.\ (1).  The neutrons in the 
proposed experiment are in fact acted on by forces having non-vanishing 
components in the direction of the beam.
\\
\\
\noindent
{\bf 2.  FORCES ON THE NEUTRON}
\\
\\
\noindent
In reality, any forces on the neutrons are exerted in the neutrons' 
exchange of momentum with atoms in the barrier.  Allman {\it et al.} 
substitute a boundary condition for the interaction of the neutrons 
with the atoms in the barrier.  That approximation gives an adequate 
wave function and leads to the correct phase shift.  No potential 
gradient appears in the Schroedinger equation, but Allman {\it et al.}, 
in saying that no force is exerted, neglect the force exerted on the 
neutrons at the boundary.  

To focus on the principle involved, consider an idealized situation in 
which a single neutron, initially in a state represented by some wave packet  
$\psi (t)$, is aimed at a long channel  
so that there is a time interval when the neutron's wave packet is for 
practical purposes
entirely within the channel.  In the best case, the wave packet enters the 
channel with only 
minimal reflection, proceeds through the channel 
at reduced speed as described by Ref.\,1, then exits the channel, 
again with minimal reflection, and continues on with its initial speed.  
Although the reflection is minimal, the wave function unavoidably spreads 
into some diffraction region around the ends of the channel.  
There the boundary exerts a force $F_b$ in the beam direction whose 
expectation is given by 

\begin{eqnarray}
\Big< F_b \Big> _t &=& {d\over dt} \Big<p_b \Big> = i\hbar \int d^3 x
\Bigg( {\partial \psi ^* \over \partial t} \nabla _b \psi -
{\partial \psi \over \partial t} \nabla _b \psi ^* \Bigg) 
\nonumber \\
&=& \mp {\hbar ^2 \over 2m} \int d^2 x \Big| \nabla _b \psi \Big| ^2 \,,
\end{eqnarray}
where the two-dimensional integral is carried over the boundary 
segments normal to the 
beam direction, $i.e.$ over the surfaces of 
the barrier outside of the channel.  In Eq.\ (2), the minus sign 
applies to the first face of the barrier encountered by the neutron 
and the plus sign to the second face, corresponding to the neutron's 
losing momentum when it enters 
the channel and regaining that momentum when it leaves.  Ehrenfest's 
theorem guarantees that the force given in Eq.\ (2) accounts 
correctly for the reduced momentum $p'$ in the channel, correctly 
given by Allman {\it et al.} as

\begin{equation}
p' = p - \Delta p \approx p - {\pi ^2 \hbar ^2 \over 2a ^2 p} \,,
\end{equation}
where $p$ is the free-space momentum before and after the passage 
through the channel.  In the case of a neutron that misses the channel 
and is reflected from the barrier, the time integral of the force given 
in Eq.\ (2) agrees with a net momentum transfer equal to $-2p$, as it must.

That the momentum shift is brought about by the force exerted on 
the neutron at the boundary is especially clear when the wave packet 
exits the channel, because there the negligible momentum carried in 
the small reflected wave is directed oppositely to the momentum 
transferred to the neutron by the force at the boundary, so the reflected
wave cannot carry off the transferred momentum.

The same physics appears in a more physical model where the barrier 
is represented by a finite repulsive potential instead of by a boundary 
condition.  If the potential is taken to be rounded at the edge of 
the barrier, a finite force appears  in the Schroedinger equation 
in the form of the gradient of the potential.  If the potential ends 
with a step, the force becomes a delta function.  Although the details 
of the force vary with those of the model, the momentum the force imparts to 
the neutron is in all cases the same as that imparted by the boundary 
in the boundary-condition model.
\\
\\
\noindent
{\bf 3.  CONCLUSIONS}
\\
\\
\noindent
This phenomenon stands in contrast to genuinely force-free 
interference phenomena [3,4,5], in which the phase shift is independent 
of the energy.  It is also very different from the Aharonov-Bohm 
magnetic scattering effect, in which electrons are scattered from a 
solenoid containing a magnetic flux.  There too the interaction of 
the beam particles with the atoms of the solenoid can adequately be 
represented by a boundary condition and the force at the boundary 
accounts correctly for the momentum change when the electrons are 
scattered [6].  The interest in that version of the Aharonov-Bohm 
effect arises from the fact that the scattering, and with it the 
momentum transfer, depends upon the magnitude of the magnetic flux in 
the solenoid even though there is no force and no momentum transfer
between the electrons and 
the local, present magnetic field.  As expected, the magnetic-flux-dependent
part of the phase shift is independent of the electron's energy.
No analogous consideration arises 
when neutrons pass through a channel in a barrier.

These considerations apply equally to the optical interference experiments in Ref.\ [1] 
as to the proposed neutron experiment, but they do not apply to the temporally modulated
constriction in a new experiment proposed by Allman {\it et al.} [7], which
involves no force in the direction of motion of the wave packet.  That proposed experiment falls
into the force-free class that includes the Aharonov-Casher effect, the Scalar Aharonov-Bohm
effect, and the force-free nuclear phase shifter but  not the electric and magnetic
Aharonov-Bohm effects [4,8].
\\
\\
\noindent
{\bf Acknowledgements}.  This work is supported by the U.S. Department of Energy, Nuclear
Physics Division, under contract W-31-109-ENG-38.  I thank Brendan Allman for informing me about the new proposed experiment.
\\
\noindent
\\
\\
{\bf REFERENCES}
\begin{enumerate}
\tighten
\item B. E. Allman, A. Cimmino, S. L. Griffin, and A. G. Klein, Found.\
Phys.\ {\bf 29}, 325 (1999).
\item J.-M. Levy-Leblond, Phys.\ Lett.\ A {\bf 125}, 441 (1987).
\item A. Zeilinger, in {\it Fundamental Aspects of Quantum Theory}, V. Gorini
and A. Frigerio, eds., (NATO ASI Series B, Vol. 144), (Plenum, NY 1986), p.\ 311.
\item M. Peshkin, Found.\ Phys.\ {\bf 29}, 481 (1999) and quant-ph/9806055
\item P. Pfeifer, Phys. Rev. Lett. {\bf 72}, 305 (1994).
\item M. Peshkin and A. Tonomura, {\it The Aharonov-Bohm Effect}, (Lecture
Notes in Physics 340), (Springer-Verlag, NY 1989), p.\ 31.
\item B. E. Allman, A. Cimmino, and A. G. Klein, Found.\ Phys.\ Lett. (to be published).  
Also see D. M. Greenberger, Physica B {\bf 151}, 374 (1988).
\item M. Peshkin and H. J. Lipkin, Phys.\ Rev.\ Lett.\ {\bf 74}, 2847 (1995).
\end{enumerate}

\end{document}